\documentclass{pasj00}
\usepackage{bm}

\DeclareAbbreviation\AAHam{Astron. Abh. Hamburg. Sternw.}
\DeclareAbbreviation\AARv{Astron. Astrophys. Rev.}
\DeclareAbbreviation\AAS{American Astron. Soc. Meeting Abstracts}
\DeclareAbbreviation\AcA{Acta Astron.}
\DeclareAbbreviation\actaa{Acta Astron.}
\DeclareAbbreviation\Afz{Astrofizika}
\DeclareAbbreviation\AGAb{Astronomische Gesellschaft Abstract Ser.}
\DeclareAbbreviation\an{Astron. Nachr.}
\DeclareAbbreviation\AnAp{Annales d'Astrophysique}
\DeclareAbbreviation\AnTok{Tokyo Astron. Obs. Annals, Sec. Ser.}
\DeclareAbbreviation\Ap{Astrophysics}
\DeclareAbbreviation\ARep{Astron. Rep.}
\DeclareAbbreviation\ATel{Astron. Telegram}
\DeclareAbbreviation\ATsir{Astron. Tsirk.}
\DeclareAbbreviation\AcApS{Acta Astrophys. Sinica}
\DeclareAbbreviation\AstL{Astron. Lett.}
\DeclareAbbreviation\BaltA{Baltic Astron.}
\DeclareAbbreviation\BANS{Bull. of the Astron. Institutes of the Netherlands Suppl. Ser.}
\DeclareAbbreviation\BASI{Bull. Astron. Soc. India}
\DeclareAbbreviation\BeSN{Be Newslett.}
\DeclareAbbreviation\BHarO{Harvard Coll. Obs. Bull.}
\DeclareAbbreviation\CBET{Cent. Bur. Electron. Telegrams}
\DeclareAbbreviation\ChJAA{Chinese J. of Astron. and Astrophys.}
\DeclareAbbreviation\caa{Chinese J. of Astron. and Astrophys.}
\DeclareAbbreviation\CoAsi{Asiago Contr.}
\DeclareAbbreviation\CoSka{Contributions of the Astronomical Observatory Skalnat\'e Pleso}
\DeclareAbbreviation\GCN{GRB Coord. Netw. Circ.}
\DeclareAbbreviation\ErgAN{Erg. Astron. Nachr.}
\DeclareAbbreviation\ibvs{IBVS}
\DeclareAbbreviation\IEEEP{IEEE Proc.}
\DeclareAbbreviation\JAD{J. Astron. Data}
\DeclareAbbreviation\JAVSO{J. American Assoc. Variable Star Obs.}
\DeclareAbbreviation\JBAA{J. Br. Astron. Assoc.}
\DeclareAbbreviation\JPhCS{J. of Physics Conference Series}
\DeclareAbbreviation\JPSJ{J. Phys. Soc. Japan}
\DeclareAbbreviation\JSARA{J. of the Southeastern Assoc. for Research in Astron.}
\DeclareAbbreviation\LowOB{Lowell Obs. Bull.}
\DeclareAbbreviation\MitAG{Mitteil. der Astronom. Gesell. Hamburg}
\DeclareAbbreviation\MitVS{Mitteil. Ver\"{a}nderl. Sterne}
\DeclareAbbreviation\MmSAI{Mem. Soc. Astron. Ital.}
\DeclareAbbreviation\Msngr{Messenger}
\DeclareAbbreviation\NewA{New Astron.}
\DeclareAbbreviation\na{New Astron.}
\DeclareAbbreviation\NewAR{New Astron. Rev.}
\DeclareAbbreviation\NInfo{Nauchnye Informatsii}
\DeclareAbbreviation\OAP{Odessa Astron. Publ.}
\DeclareAbbreviation\Obs{Observatory}
\DeclareAbbreviation\OEJV{Open Eur. J. on Variable Stars}
\DeclareAbbreviation\PASA{Publ. Astron. Soc. Australia}
\DeclareAbbreviation\PASAu{Publ. Astron. Soc. Australia}
\DeclareAbbreviation\PCCP{Phys. Chem. Chem. Phys.}
\DeclareAbbreviation\PAZh{Pis'ma AZh}
\DeclareAbbreviation\PhR{Phys. Rep.}
\DeclareAbbreviation\PVSS{Publ. Variable Stars Sect. R. Astron. Soc. New Zealand}
\DeclareAbbreviation\PZ{Perem. Zvezdy}
\DeclareAbbreviation\PZP{Perem. Zvezdy, Prilozh.}
\DeclareAbbreviation\QJRAS{QJRAS}
\DeclareAbbreviation\RA{Ricerche Astronomiche}
\DeclareAbbreviation\RMxAA{Rev. Mexicana Astron. Astrof.}
\DeclareAbbreviation\RvMA{Reviews of Modern Astron.}
\DeclareAbbreviation\SASS{Society for Astronom. Sciences Ann. Symp.}
\DeclareAbbreviation\Sci{Science}
\DeclareAbbreviation\SPIE{SPIE Proc.}
\DeclareAbbreviation\SvA{Soviet Astronomy}
\DeclareAbbreviation\SvAL{Soviet Astronomy Letters}
\DeclareAbbreviation\VeSon{Ver\"{o}ff. Sternw. Sonneberg}
\DeclareAbbreviation\VSOLJBul{VSOLJ Variable Star Bull.}
\DeclareAbbreviation\yCat{VizieR Online Data Catalog}
\DeclareAbbreviation\ZA{Z. Astrophys.}

\def\PublisherSpringer{Berlin: Springer-Verlag}

\begin{document}
\SetRunningHead{T. Kato and M. Uemura}{Period Analysis using Lasso}

\Received{201X/XX/XX}
\Accepted{201X/XX/XX}

\title{Period Analysis using the Least Absolute Shrinkage and Selection Operator (Lasso)}

\author{Taichi \textsc{Kato}}
\affil{Department of Astronomy, Kyoto University,
       Sakyo-ku, Kyoto 606-8502}
\email{tkato@kusastro.kyoto-u.ac.jp}

\and

\author{Makoto \textsc{Uemura}}
\affil{Astrophysical Science Center, Hiroshima University, Kagamiyama, 1-3-1
       Higashi-Hiroshima 739-8526}


\KeyWords{
         methods: data analysis
          --- methods: statistical
          --- stars: individual ($\delta$ Cep, R Sct)
         }

\maketitle

\begin{abstract}
   We introduced least absolute shrinkage and selection operator
(lasso) in obtaining periodic signals in unevenly spaced time-series data.
A very simple formulation with a combination of a large set of sine and
cosine functions has been shown to yield a very robust estimate,
and the peaks in the resultant power spectra were very sharp.
We studied the response of lasso to low signal-to-noise data,
asymmetric signals and very closely separated multiple signals.
When the length of the observation is sufficiently long, all of them
were not serious obstacles to lasso.  We analyzed the 100-year visual
observations of $\delta$ Cep, and obtained a very accurate period
of 5.366326(16) d.  The error in period estimation was several times
smaller than in Phase Dispersion Minimization.
We also modeled the historical data of R Sct, and obtained a reasonable
fit to the data.  The model, however, lost its predictive ability after
the end of the interval used for modeling, which is probably a result
of chaotic nature of the pulsations of this star.
We also provide a sample R code for making this analysis.
\end{abstract}

\section{Introduction}

   There has been a wealth of history in analyzing periodicities
in time-series data, particularly for variable stars.
Determination of periods in variable stars is a very fundamental
issue in that they are used in many applications, including
classification of variable stars, calibration of the period-luminosity
relations, determination of the pulsation modes, detection of stellar
rotation, detection of exoplanet transits and determination of
the precession mode and frequency in accretion disks.
These time-series data in practical astronomy are not usually
sampled evenly in the time domain, and it is difficult to apply
fast Fourier transform (FFT) directly as in other fields of
science.  Alternatively, there have been well-established methods
based on discrete Fourier transform \citep{DeemingDFT}
and on least-squares fit of sinusoids for unevenly spaced data
(Lomb-Scargle Periodogram, \cite{LombScargle}; \cite{hor86periodanalysis};
\cite{zec09LombScargle}).
This type of method has been extended to extract
finite numbers of signals by subsequently subtracting the strongest
signals (CLEAN; \cite{CLEAN}).

   Another class of approach has been taken by evaluating dispersions
either in sum of lengths between phase-sorted data
\citep{dwo83stringlength} or sum of dispersions in phased bins
against trial periods.  This class of approach includes
\citet{LaflerKinman}, well-used Phase Dispersion Minimization
(PDM, \cite{PDM}) and Analysis of Variance (AoV, \cite{sch89AoV}).

   Yet another class of approach to estimate the power spectrum is
by assuming a sum of $\delta$-function-like poles on a frequency
space extended to the full complex space (\cite{ModernSpectrumAnalysis},
chap. 2; \cite{kay81spectrumanalysis}).  This approach is generally
referred to as Maximum Entropy Method (MEM) or autoregressive model
(AR).  Although this approach is potentially very powerful in detecting
a small number of sharp signals in the power spectrum, the usage in
practical astronomical data has been limited because the well-known
quick algorithm (Burg algorithm) using a recursive technique can
only be applied to evenly spaced data \citep{bur67MEM}.

   In recent years, there has been a remarkable progress in the
field of Compressed Sensing, which deals with a class of problems
in restoring or estimating sparsely scattered, finite number
of parameters in a huge dimension.  We present an application of
this method to period analysis.

\section{Methods}

\subsection{Decomposition to Fourier Components}

   The following description basically follows the mathematical
formulation reviewed in \citet{tan10compressedsensing}.
[For a review of Compressed Sensing, see \citet{don06compressedsensing}].
We assume a time-series data $Y(t_i)$ with unevenly spaced
times of observations at $t_i$.  The data are assumed to be subtracted
for their average, and the mean of $Y$ is zero.
The observation can be expressed as a sum of signal ($Y_s$) and
random errors ($n$) :
  \begin{equation}\label{equ:obserror}
  Y_i = Y(t_i) = Y_s(t_i) + n(t_i).
  \end{equation}
We assume that the signal is composed of a sum of strictly periodic
functions.
Using sine and cosine Fourier components, $Y_s$ can be expressed as :
  \begin{equation}\label{equ:fouriersum}
  Y_s(t_i) = \sum_j{a_j \cos (\omega_j t_i)} + \sum_j{b_j \sin (\omega_j t_i)},
  \end{equation}
where $\omega$ are frequencies and $a$ and $b$ are amplitudes.
The problem can be set to estimate $a$ and $b$ from $Y$.
If the number of different $\omega$ is larger than the half number of
observations, the equation cannot be solved uniquely.
Rewriting
  \begin{equation}
  \bm{y} = (Y_1, \cdots, Y_N)^{\mathrm{T}}
  \end{equation}
and
  \begin{equation}
  \bm{x} = (a_1, \cdots, a_M, b_1, \cdots, b_M)^{\mathrm{T}},
  \end{equation}
where $N$ and $M$ are number of observations and number of
different $\omega$, respectively.
We can rewrite a set of equations \ref{equ:obserror} and
\ref{equ:fouriersum} as a form of
  \begin{equation}
  \bm{y} = A\bm{x} + \bm{n}
  \end{equation}
using a $2M \times N$ observation matrix $A$ composed of
  \begin{equation}\label{equ:fouriercomp}
  A_{i,j} = \left\{
    \begin{array}{ll}
      \cos (\omega_i t_j), & \mbox{($i \le M$)} \\
      \sin (\omega_{i-M} t_j), & \mbox{($i > M$)}.
    \end{array}
    \right.
  \end{equation}
We define $\bm{x}_0$ as the vector to be estimated.

   The problem is to estimate sparse $\bm{x}_0$, i.e., with
only small number of non-zero elements, under the restriction of
$\bm{y} = A\bm{x}$, if the time-series data is expected to have 
only a few periodic components, which is often the case with 
period analysis.  We define ``0-norm'' $||\bm{x}||_0$
as the number of non-zero elements in $\bm{x}_0$.  The problem
can be then formalized to choose $\bm{x}_0$: 
  \begin{equation}
  \hat{\bm{x}}^{(0)} = \arg \min_{\bm{x}}||\bm{x}||_0 \quad \mathrm{subj.\; to} \quad \bm{y} = A\bm{x}.
  \end{equation}
This estimate is called $\ell_0$ regularization, and
is known to be very robust in this problem setting.
The problem, however, is that $\ell_0$ regularization
is computationally difficult, and is known to be an NP-hard
problem \citep{nat95sparse}.

  In order to overcome this difficulty, the following alternative
$p$-norms are usually employed: 
  \begin{equation}
  ||\bm{x}||_p = \left\{
    \begin{array}{ll}
      (\sum_{i=1}^N|x_i|^p)^{(1/p)}, & \mbox{($p > 0$)} \\
      ||\bm{x}||_0, & \mbox{($p = 0$)}. \\
    \end{array}
    \right.
  \end{equation}
In particular, $p=1$ ($\ell_1$) regularization has
an advantage in that it can be solved by conventional ways:
by introducing a vector $\bm{u} = (u_1, \cdots, u_N)^{\mathrm{T}}$
composed of non-negative elements, the equation becomes
  \begin{equation}
  \min \sum_{i=1}^Nu_i \quad \mathrm{subj.\;to} \quad -\bm{u} \le \bm{x} \le \bm{u},\quad \bm{y} = A\bm{x},
  \end{equation}
and it is a well-known linear programming (LP).
Most notably, it has been recently demonstrated that this $\ell_1$
regularization is also the sparsest solution ($\ell_0$
regularization) in most large underdetermined systems
(see e.g., \cite{don06compressedsensing}; \cite{can06signalrecovery}).

   There have been a number of methods using this $\ell_1$
constrained estimations.  Among them, \citet{lasso} presented
a concept of least absolute shrinkage and selection operator
(lasso).  This estimation has become well-known after the development
of a fast algorithm, known as least angle regression (LAR) by \citet{LARS}.
The lasso is defined as:\footnote{
   In actual calculation of lasso estimates, we need to normalize
   parameters, i.e., $\sum_i{A_{i,j}} = 0$, $\frac{1}{N} \sum_i{A_{i,j}^2} = 1$, 
   for $j = 1, \cdots, 2M$ \citep{glmnet}.
   We here introduce a conceptual formula and avoided complicated
   use of normalization factors.  The {\bf glmnet} package in R
   automatically performs this normalization and we usually do not
   need to consider the normalization.
}
  \begin{equation}\label{equ:lassodef}
  \hat{\bm{x}}^{\mathrm{LAR}} = \arg \min_{\bm{x}} (F(\bm{x}) \equiv \frac{1}{2N} ||\bm{y} - A\bm{x}||_2^2 + \lambda ||\bm{x}||_1),
  \end{equation}
where $\lambda ||\bm{x}||_1$ is the $\ell_1$-norm penalty function
with a parameter $\lambda$ having a value of $\lambda \ge 0$.
This estimate becomes identical with a least-squares estimation
at $\lambda = 0$.

   The optimal value of $\lambda$ may be chosen by using the
cross-validation technique:
after breaking the data into two groups,
use the one group for lasso analysis and use the rest of data
for evaluation of residuals.  Mean squared errors (MSE)
were evaluated using several randomly chosen different
partitioning of data.  One can then estimate
$\lambda$ giving the smallest MSE.  In application to analysis of
actual time-series data, it might be helpful to survey the response
of lasso using different $\lambda$ values, rather than fixing at
$\lambda$ giving the smallest MSE, since the derived periods are
not very different between different $\lambda$ values, and since
smaller $\lambda$ gives smaller number of false signals, which
is profitable when it is already known that only small number of
signals are present in the data.  An example of cross-validation
diagram is shown in subsection \ref{sec:deltacep}
(figure \ref{fig:cepdeltacv}).

   We must note that the present application of $\ell_1$ regularization
using sinusoidal functions is not unprecedented in astronomy,
but that there was at least one in the pre-LAR era
in analyzing radial-velocity data \citep{che98specanalysis}
based on the same scheme.  We introduce the present method because
time-series data in variable stars are commonly met in practical
astronomy and are very suitable targets for the present analysis,
either in detection of multiple periods or precise determination of
the periods.  Furthermore, the development of LAR algorithm and
its public availability on a popular platform has enabled 
an application of the Compressed Sensing far easier to reach 
than in the time of \citet{che98specanalysis}.

\subsection{Numerical Solution}

   This LAR algorithm has been implemented as the package {\bf lars}
in R software\footnote{
   The R Foundation for Statistical Computing:\\
   $<$http://cran.r-project.org/$>$.
} and we used this package combined with {\bf glmnet}
(generalized linear model via penalized maximum likelihood)
as a wrapper, which also provides cross-validation results.
See appendix for a sample code.

\section{Results}

\subsection{Response to Artificial Data: Single Frequency}

\begin{figure}
  \begin{center}
    \FigureFile(88mm,50mm){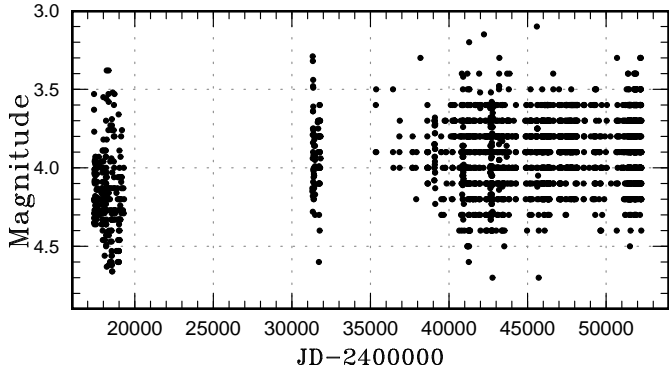}
  \end{center}
  \caption{VSOLJ observations of $\delta$ Cep.}
  \label{fig:deltacep}
\end{figure}

\begin{figure}
  \begin{center}
    \FigureFile(60mm,120mm){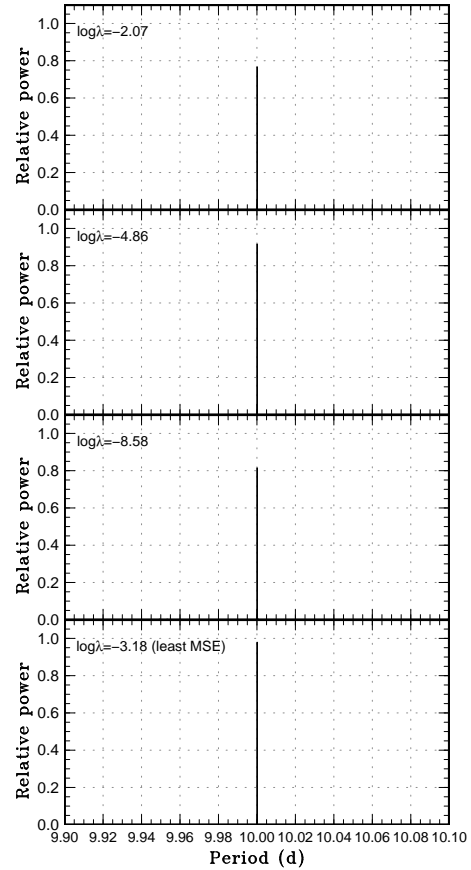}
  \end{center}
  \caption{Analysis of a pure sine wave with noises.}
  \label{fig:artdata}
\end{figure}

\begin{figure}
  \begin{center}
    \FigureFile(60mm,120mm){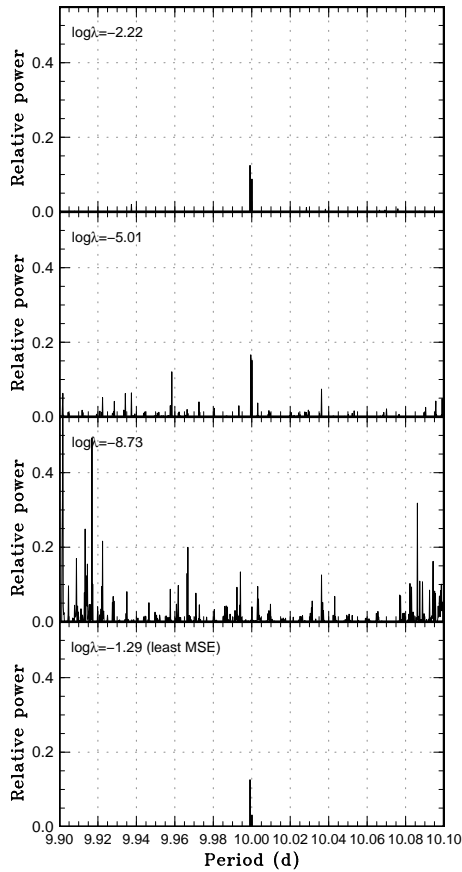}
  \end{center}
  \caption{Analysis of a pure sine wave with noises five times larger
  than the signal.}
  \label{fig:artdata2}
\end{figure}

\begin{figure}
  \begin{center}
    \FigureFile(60mm,120mm){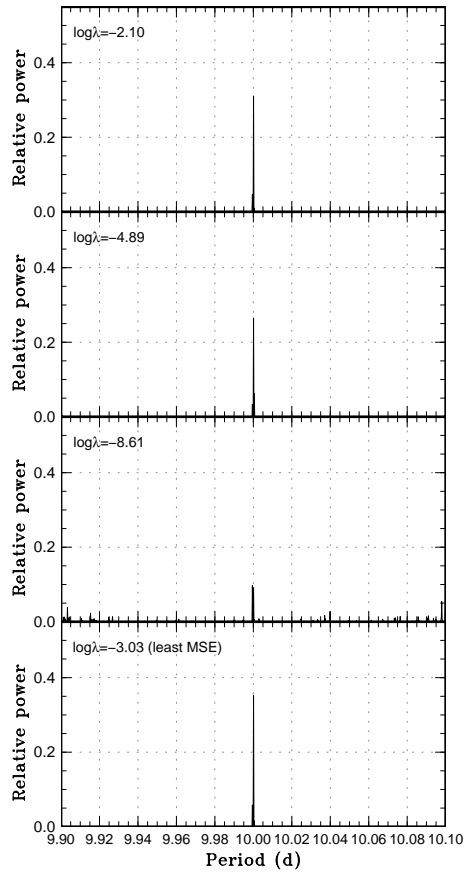}
  \end{center}
  \caption{Analysis of non-sinusoidal wave.}
  \label{fig:artdata3}
\end{figure}

   We examined the response of this analysis to the artificial data.
In order to reflect typical astronomical time-series observations,
we used observations of $\delta$ Cep by Variable Star Observers' League
in Japan (VSOLJ), spanning 1906 through 2001, with a total
number of 2955 visual estimates having large gaps (1911--1944, 1946--1955)
and inhomogeneous density of observations (figure \ref{fig:deltacep}).
We presented response to different $\lambda$ values\footnote{
   In referring to $\log \lambda$, we used natural logarithms
   throughout the paper.
}
as well as the model at the smallest MSE.
We used the times of these observations and replaced the magnitudes with
artificial data.  The first one (figure \ref{fig:artdata}) is
a pure sine wave with a period of 10 d with Gaussian noises whose
$\sigma$ is the same as the amplitude of the wave.
In this case, the true period was always reproduced within errors
of $10^{-4}$ d.  Only small noises appear when $\lambda$ is very small.
The second case is with Gaussian noises five times larger than
the amplitude of the wave (figure \ref{fig:artdata2}).
In this case, the signal was completely lost when $\lambda$ is small
(over-expression of the noises by Fourier components).
This result indicates that we need to choose adequate $\lambda$
depending on the signal-to-noise ratio of the data.  The model with
the smallest MSE is found to be an adequate solution in this case.
The third case is the asymmetric waveform.  We added second
overtone (a period of 5 d) having a half amplitude of the primary
wave.  The noises were as in the first case (figure \ref{fig:artdata3}). 
The estimates of the period were observed to slightly deviate from
the correct value (typically 2 10$^{-4}$ d), and the deviation depends
on $\lambda$.
\begin{figure}
  \begin{center}
    \FigureFile(60mm,120mm){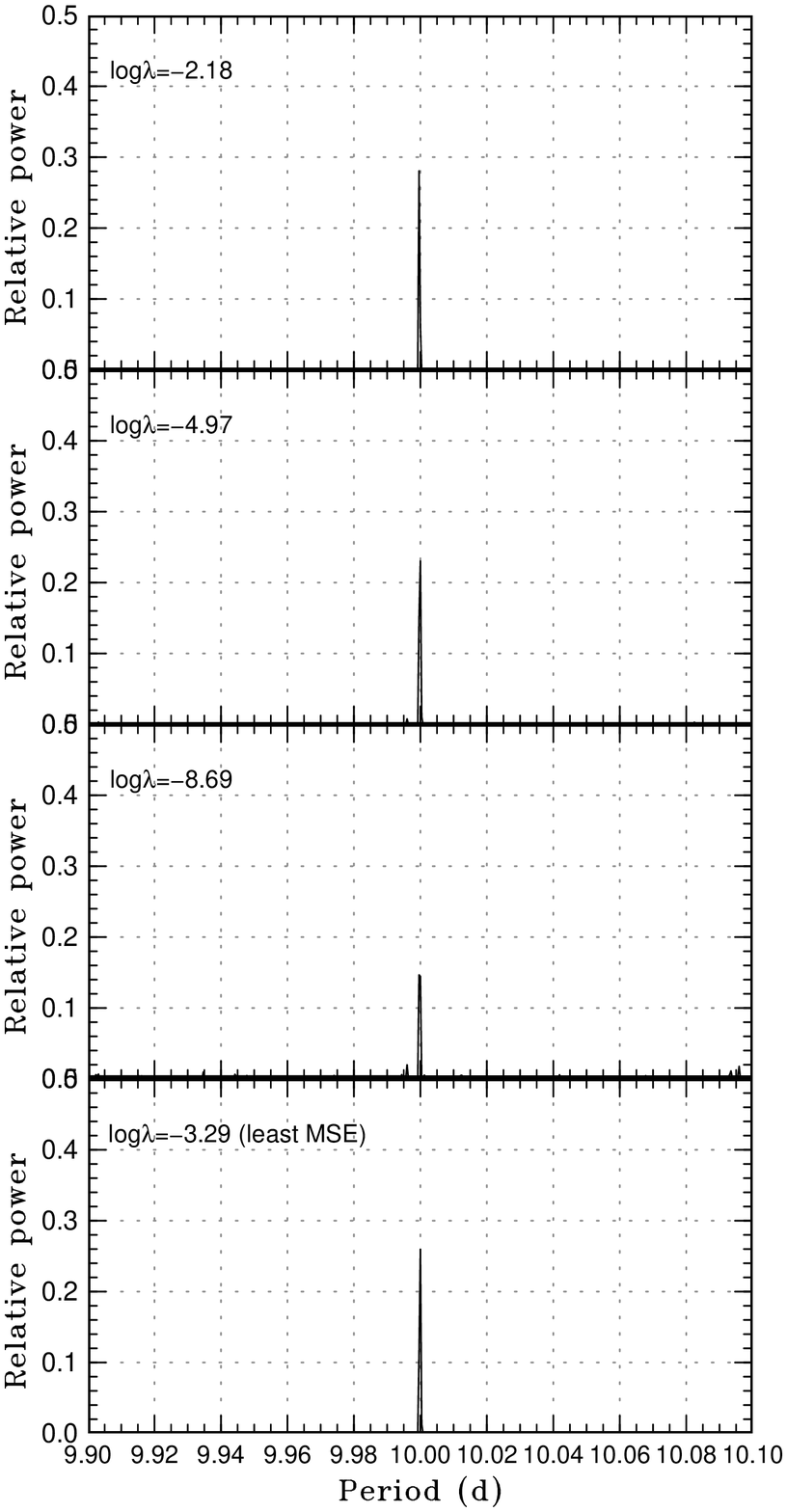}
  \end{center}
  \caption{Analysis of a non-sinusoidal wave, fitted with
  a second spectral window.}
  \label{fig:artdata4}
\end{figure}
We allowed two spectral windows 9.9--10.1 d and 4.95--5.05 d and
obtained figure \ref{fig:artdata4}.  After allowing the second
window, the main period was more clearly detected.  If cycle numbers
are sufficient as in this case, asymmetric profiles did not
significantly affect the results.

\subsection{Separation of Peaks}

\begin{figure}
  \begin{center}
    \FigureFile(60mm,120mm){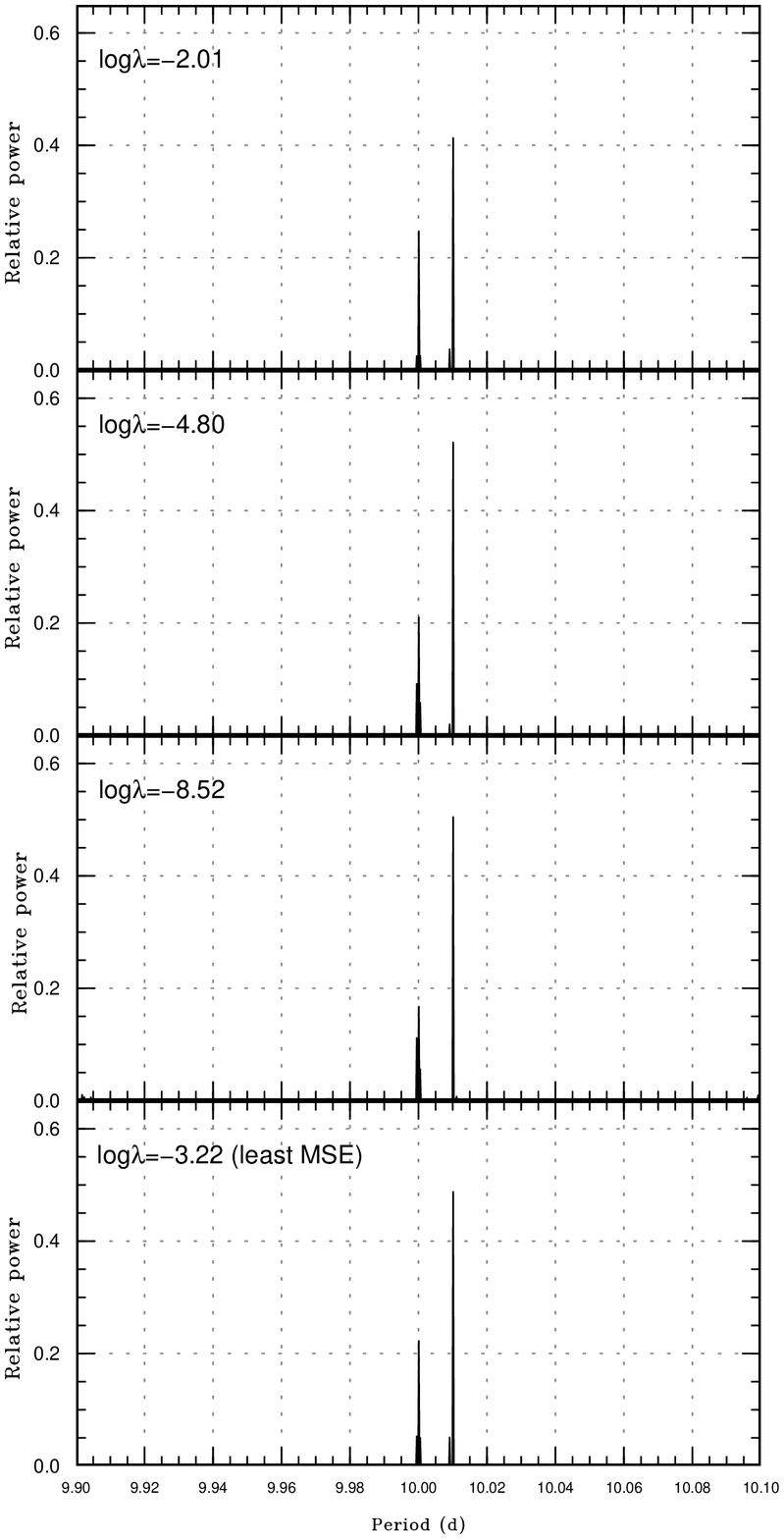}
  \end{center}
  \caption{Analysis of closely separated double waves.}
  \label{fig:artdata5}
\end{figure}

   We then examined the ability of this lasso estimate for
identifying very closely separated signals.
We analyzed a combination of two periods of 10 d and 10.005 d
and they were always detected as separate signals regardless of
the different sequence of random numbers used to add noises 
(figure \ref{fig:artdata5}).
Assuming an effective baseline of
15000 d, these two periods correspond to a difference of 0.75 cycle
between the limits of the baseline.
In a case of 10 d and 10.001 d,
we could resolve the signals in about half trials.  It was impossible
to resolve a combination of 10 d and 10.0005 d.  This degree of
resolution is, naturally, far beyond the reach of conventional
techniques, such as PDM and Fourier-type analysis.

\subsection{Analysis of $\delta$ Cephei}\label{sec:deltacep}

\begin{figure}
  \begin{center}
    \FigureFile(60mm,120mm){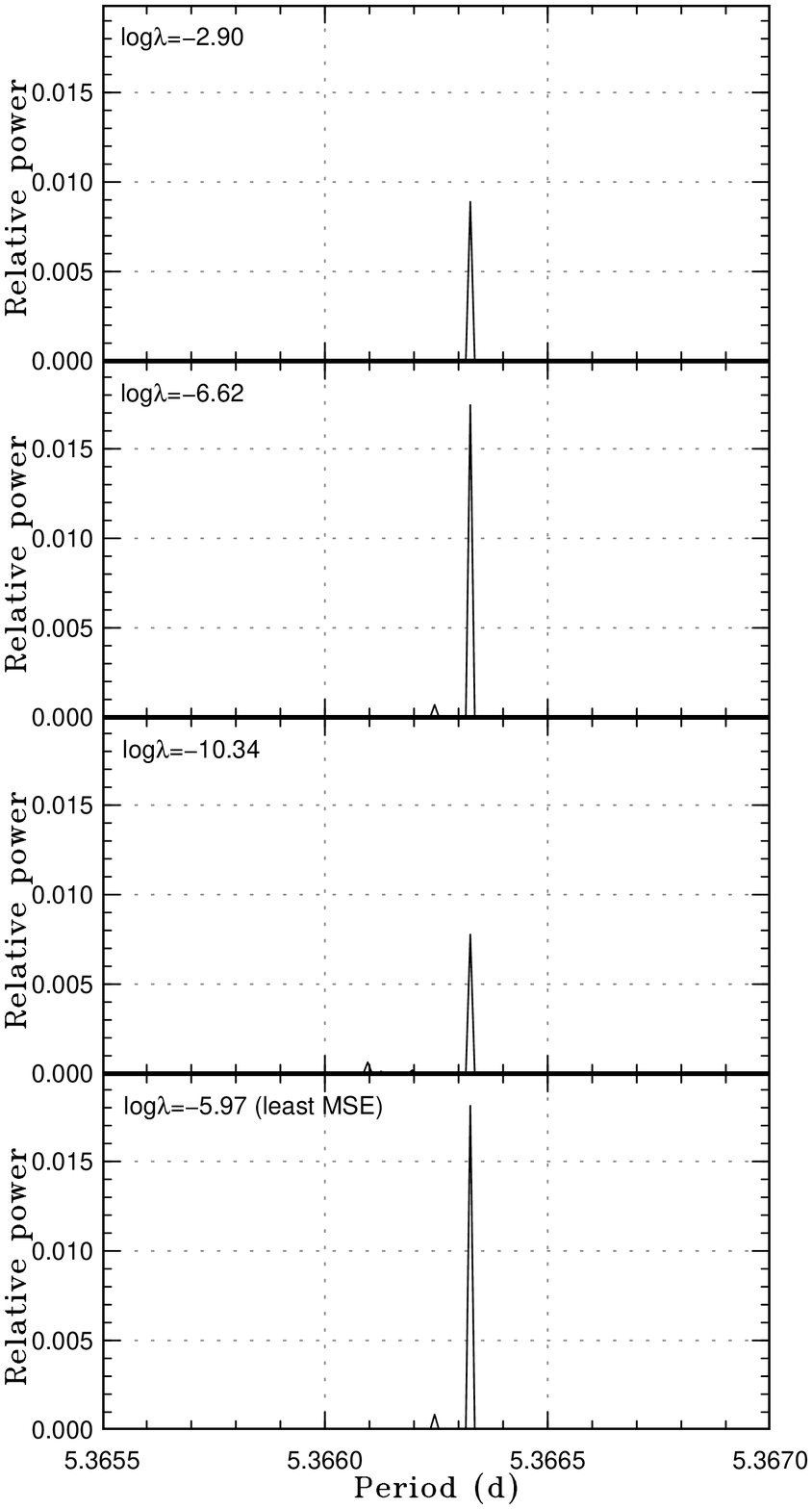}
  \end{center}
  \caption{Analysis of VSOLJ data of $\delta$ Cep.}
  \label{fig:cepdeltaper}
\end{figure}

\begin{figure}
  \begin{center}
    \FigureFile(88mm,50mm){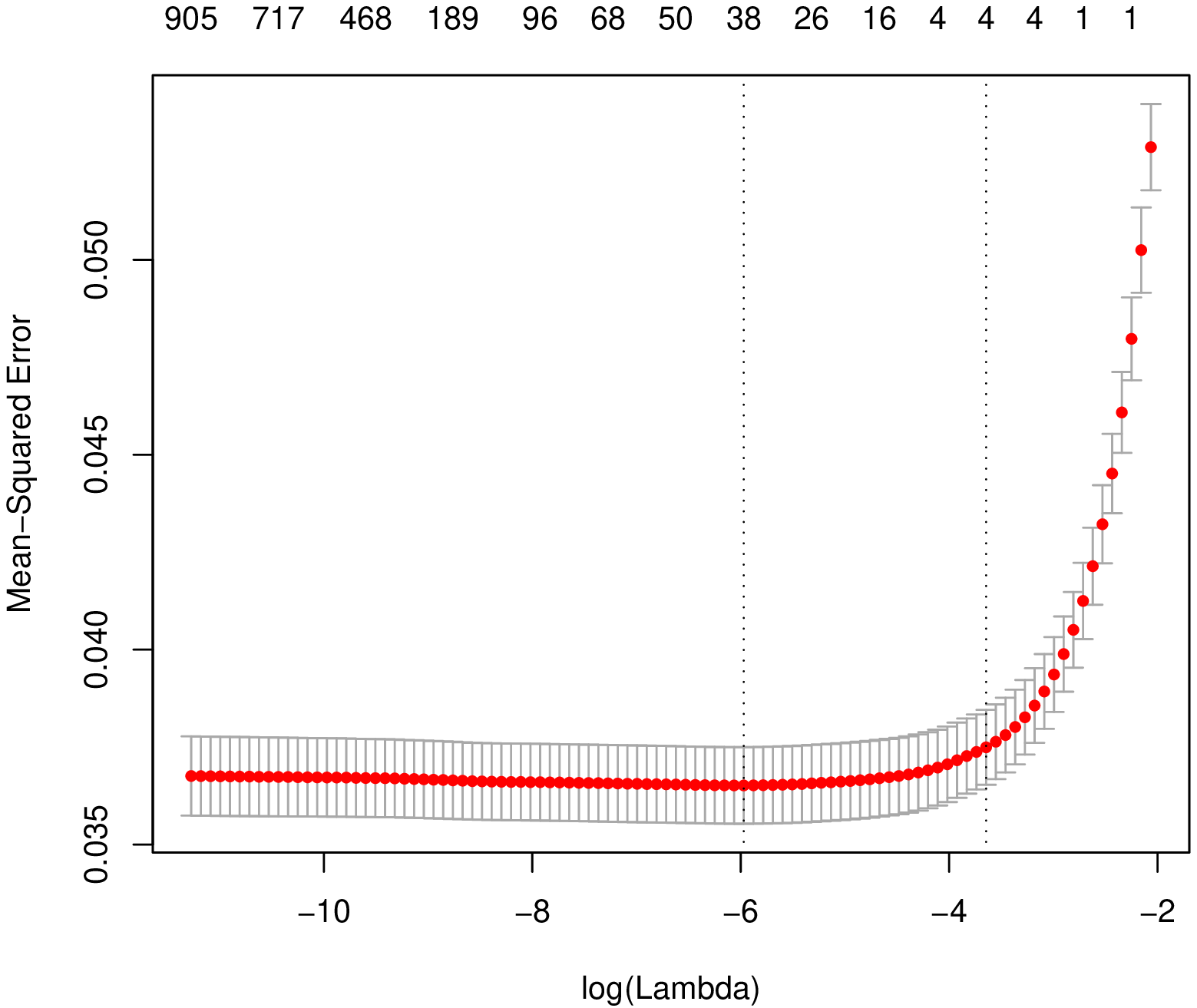}
  \end{center}
  \caption{Cross-validation diagram for $\delta$ Cep.  The points and
    error bars represent average mean-squared errors (MSEs) of
    cross-validation and their standard deviations, respectively.
    The left vertical dotted line represents the location of the
    least MSE.  The right vertical dotted line represents the most
    regularized model with a cross-validation error within one
    standard deviation of the minimum.  The numbers on the upper
    border of the figure represents numbers of non-zero coefficients
    for each value of $\lambda$.}
  \label{fig:cepdeltacv}
\end{figure}

   We performed an analysis of $\delta$ Cep itself.  The examined
periods were 1000 equally spaced bins between 5.36 and 5.37 d.
Since the peaks were very sharp, we enlarged the figure to show
the profile (figure \ref{fig:cepdeltaper}).  The cross-validation
diagram is shown in figure \ref{fig:cepdeltacv}, which indicates
that a combination of 5--40 frequencies best describes the data.
The resultant (strongest) period was 5.366326 d (in actual calculation,
we used ten times smaller bins to obtain this precision).
The period in General Catalog of Variable Stars
\citep{GCVS} is 5.366341 d.  The period obtained by PDM was
5.36629(4) d (the error was estimated by the methods of
\cite{fer89error} and \cite{Pdot2}).  Since lasso estimate is
highly nonlinear as in MEM, it is difficult to estimate the error
of the obtained period.  We alternatively performed delete-half
jackknifing with random subsampling \citep{sha95jackknifebook}:
randomly selected 50 \% of observations
and obtained periods from 10 different sets.  The resultant
1-$\sigma$ standard deviation was 0.000016 d.  The estimate of
the error appears to be consistent with the difference between
our result and the GCVS period, and lasso estimate is several
times more accurate than with the PDM in the present case.
An addition of random typical errors (0.1 mag) of visual observations
did not affect the result, indicating that lasso estimate is
very robust in detecting strictly periodic signal under the presence
of large errors.

\subsection{Application to R Scuti}

   R Sct is a well-known RV Tau-type variable star, characterized
by the presence of alternating deep and shallow minima.
Many RV Tau-type variables exhibit irregularities and the regular
alternations of minima are often disturbed.  R Sct is notable
in its irregular behavior among RV Tau stars
(e.g. \cite{kol90rsct}; \cite{buc96rsct}).
\citet{buc96rsct} reported that Fourier decomposition reasonably
well expresses the observed features of the light curve, but that
its predictive power is limited.  Since their Fourier decomposition
was not based on $\ell_1$ regularization, it would be interesting
to see whether the periods selected by $\ell_1$ regularization
equally well describe the light curve and whether the model
based on these periods has a better predictive power.

\begin{figure}
  \begin{center}
    \FigureFile(88mm,50mm){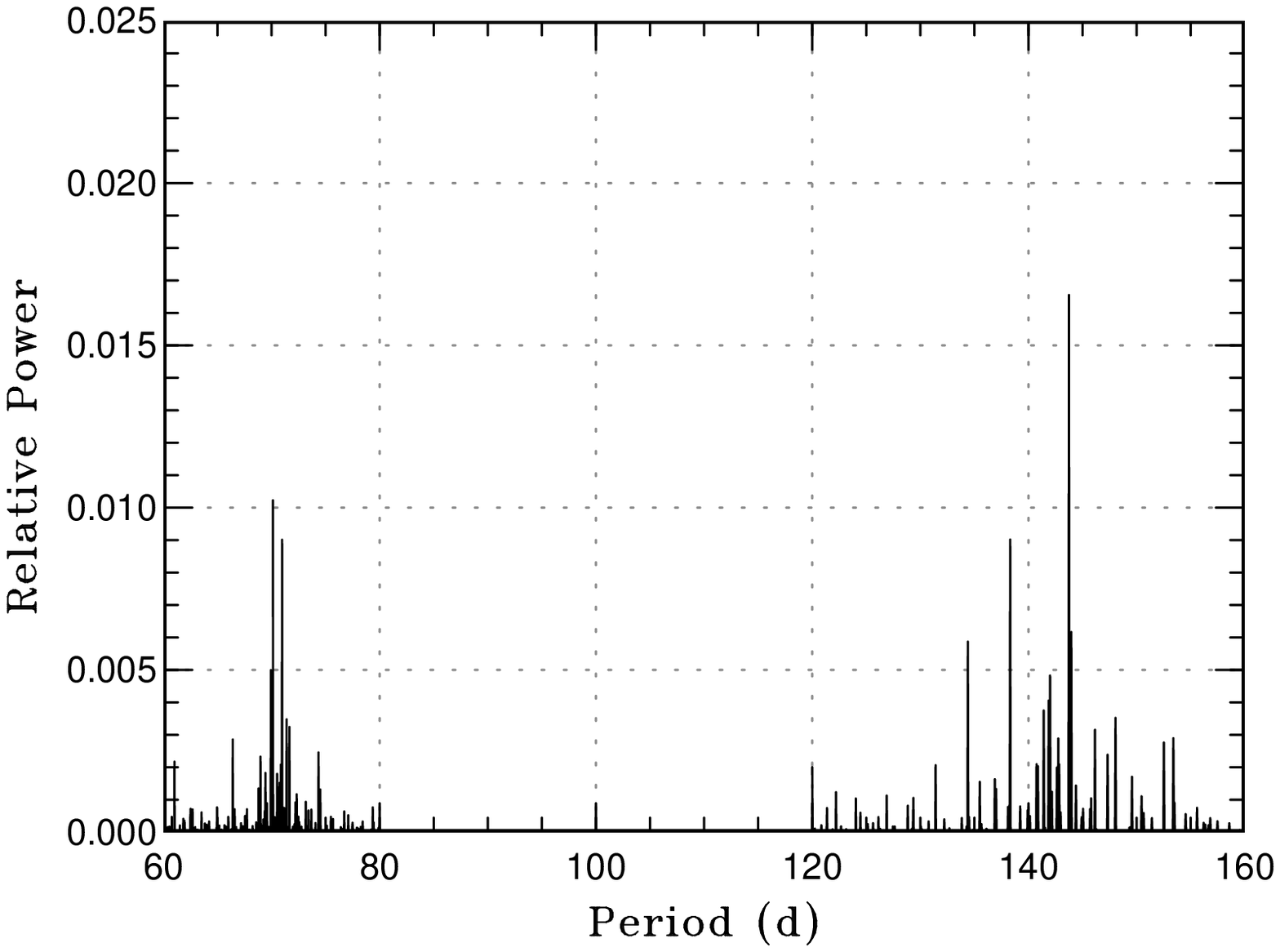}
  \end{center}
  \caption{Lasso analysis of R Sct.  We used $\log \lambda = -6.27$,
  which is the most regularized model with a cross-validation error
  within one standard deviation of the minimum.}
  \label{fig:rsctpow}
\end{figure}

   The data used were VSOLJ observations between 1906 and 2001
(16767 points).  We used two windows of periods 60--80 d and
120--160 d.  The strongest signals were at 143.8 d and 70.9 d,
associated with weaker signals (figure \ref{fig:rsctpow}).
We used $\log \lambda = -6.27$, which is the most
regularized model with a cross-validation error within one
standard deviation of the minimum.  Although the minimum MSE was
reached by $\log \lambda = -10.18$, we did not adopt it
because it was dominated by false signals and because it is physically
less likely that such a large number of pulsation modes coexist
in a radially pulsating star.
It is noteworthy that the longer period is not the twice of
the shorter period, in contrast to Fourier analysis \citep{kol90rsct}
or our own analysis with the PDM (142.1 d and 71.0 d).
These periods may represent fundamental and first-overtone periods
close to the 2:1, but slightly different, resonance which is expected
to explain the RV Tau-type phenomenon \citep{tak83rvtau}.

\begin{figure}
  \begin{center}
    \FigureFile(88mm,176mm){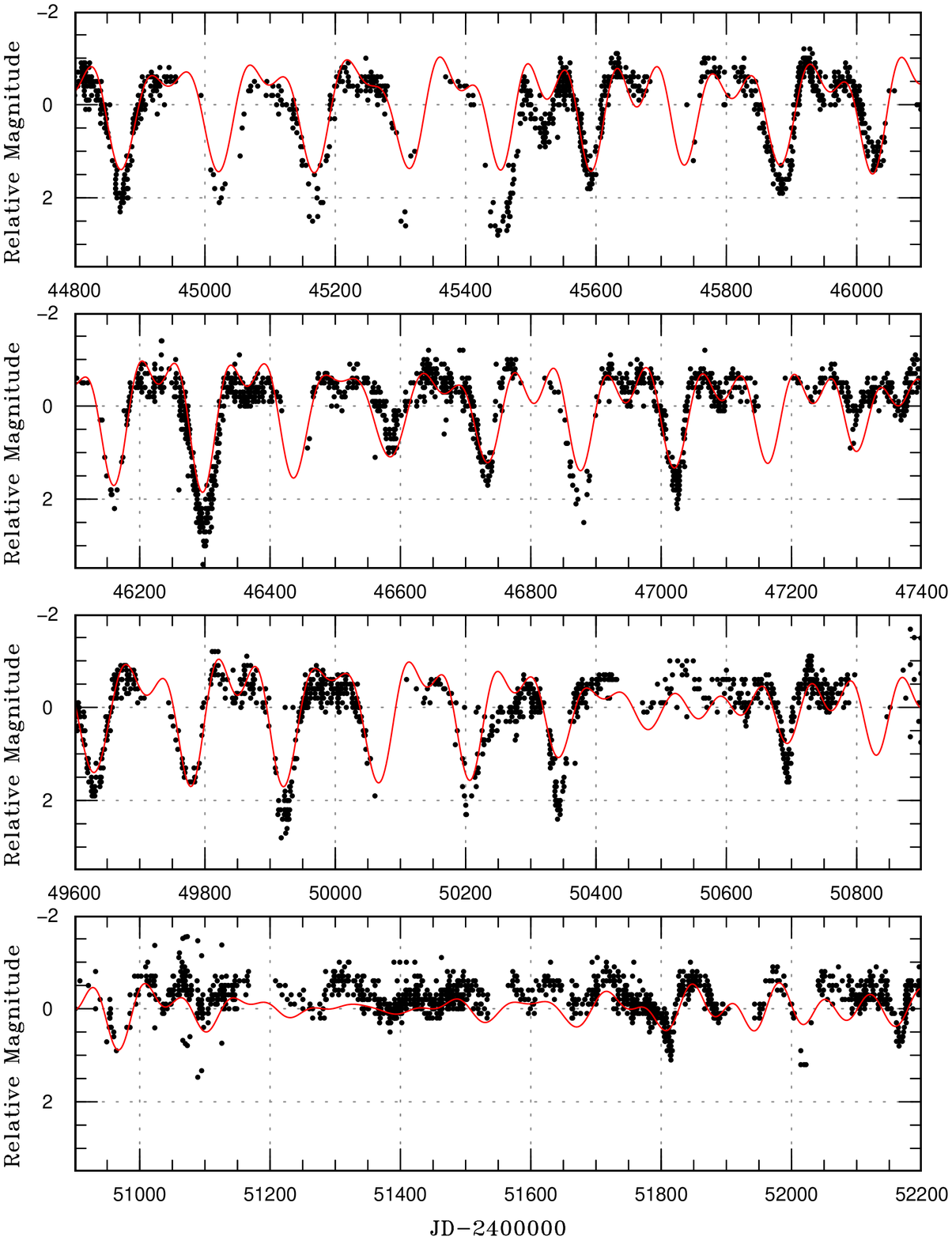}
  \end{center}
  \caption{A segment of lasso modeling to observations of R Sct.
  The points represent observations and the curves represent
  a lasso model.}
  \label{fig:rsctfit}
\end{figure}

   The lasso model seems to adequately express observations
(figure \ref{fig:rsctfit}), indicating that varying amplitudes
are basically a result of a combination of waves with periods
close to the 2:1, but slightly different, resonance.  It would be
noteworthy that interval with very low amplitudes
(e.g. JD 2451000--2451600) is well expressed by this model.

\begin{figure}
  \begin{center}
    \FigureFile(88mm,176mm){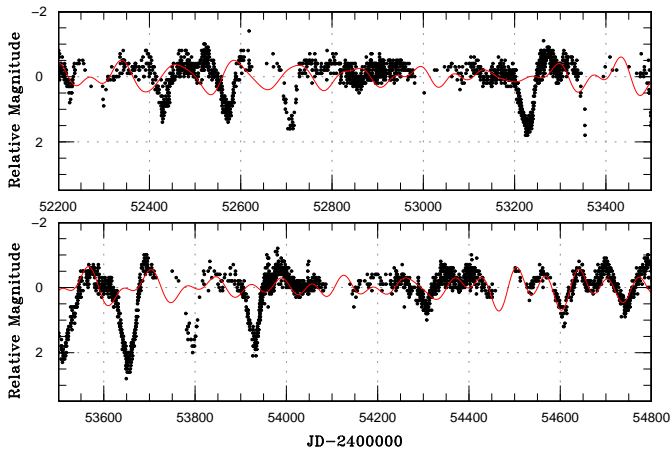}
  \end{center}
  \caption{Lasso predictions of R Sct.
  The points represent VSNET \citep{VSNET} observations and
  the curves represent a lasso model.  The model soon lost
  the predictive ability after the end of the interval (JD 2452240)
  used for modeling.}
  \label{fig:rsctexp}
\end{figure}

\begin{figure}
  \begin{center}
    \FigureFile(88mm,176mm){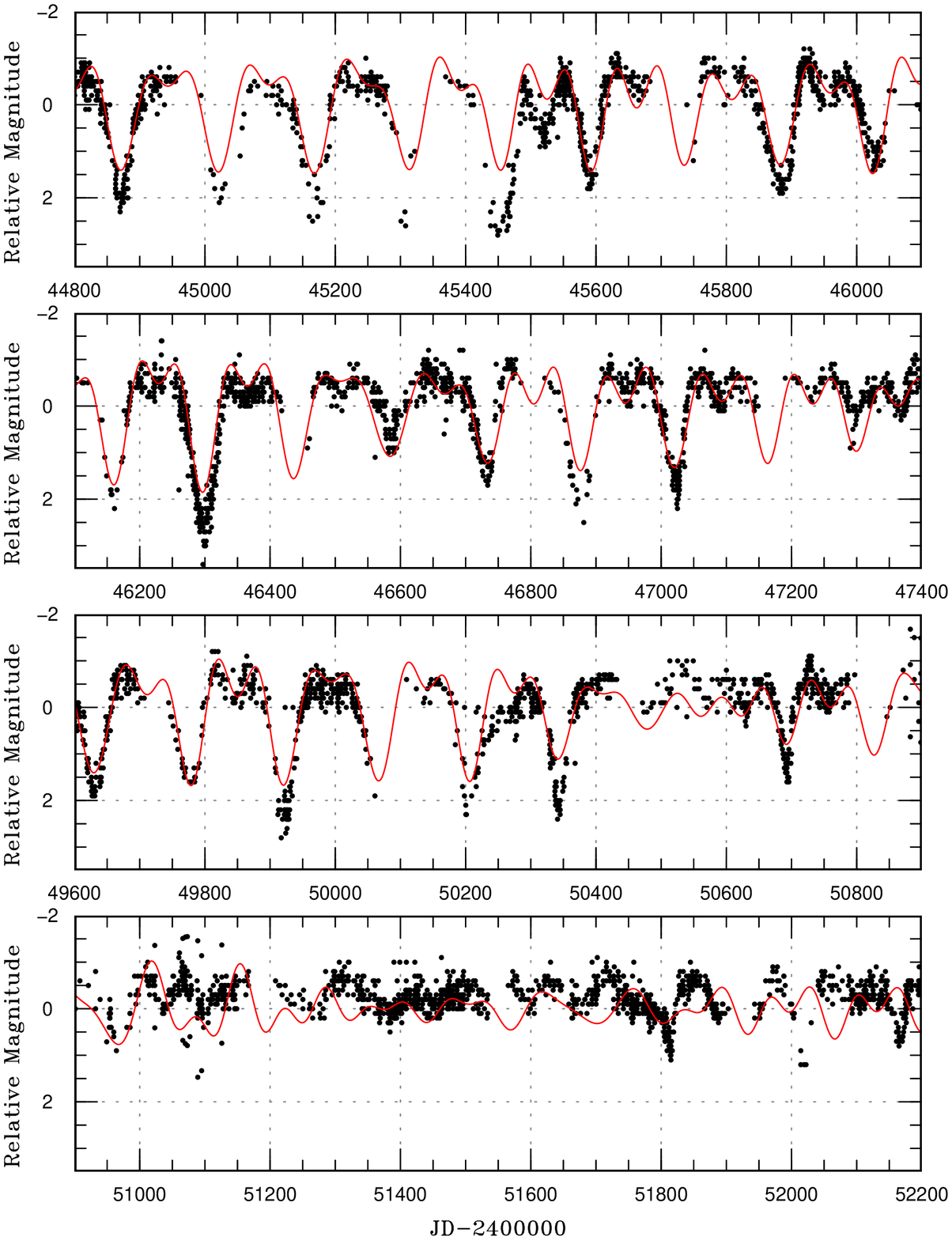}
  \end{center}
  \caption{A segment of lasso modeling to observations of R Sct.
  The data are the same as in figure \ref{fig:rsctfit}.
  The points represent observations and the curves represent
  a lasso model fitted for the data before JD 2451000.}
  \label{fig:rsctfit2}
\end{figure}

   Figure \ref{fig:rsctexp} presents an example of lasso predictions.
The model soon lost the predictive ability after the end of the epoch
(JD 2452240) used for modeling.  We also obtained a model
to the same data restricted to JD before 2451000.  Although the model
well describes the observations used in the fit, it again
loses the ability after the end of the epoch used for modeling
(figure \ref{fig:rsctfit2}).  These results are similar to that
in \citet{buc96rsct}, and it is likely that a simple combination of
multiple sinusoidal waves is not an adequate model for expressing
the behavior of this object.  This is probably caused by the chaotic
nature of the pulsations in this star.

\subsection{Splitting of Signals for Badly Phased Data}

   Although equation (\ref{equ:fouriercomp})-type bases are mathematically
sufficient for ordinary Fourier analysis, and this formulation for
lasso analysis is adequate for many cases (with sufficient number of data
and length of time-series), the lasso response to even a simple
cyclic function can yield split signals due to the lasso's
characteristics of its norm.
This is problematic because it results in erroneous estimation of
periods.  Whereas we did not observe the problem in the examples
presented so far, the problem can arise in certain applications of
lasso-based methods to period analysis, and is therefore potentially
harmful. We therefore discuss in this subsection what precisely
the problem is, and then propose a means to circumvent it.

   This situation can be understood by a simple example.
If observed data has a form of $y(t) = \sin(\omega t)$, the lasso
result has an expected form of $\hat{y}^{\mathrm{LAR}}(t) = c \sin(\omega t)$,
unless $\lambda$ is too small.  Although the amplitude $c$ is different
from 1, we can get the expected frequency $\omega$.
If the observed data, however, has a form
  \begin{equation}
  y(t) = \sin(\omega t) + \cos(\omega t), \label{equ:sinandcos}
  \end{equation}
there are different solutions with equally acceptable 
squared residuals in a form of
  \begin{equation}
  \hat{y}(t) = c_1 \sin(\omega_1 t) + c_2 \cos(\omega_2 t),
  \end{equation}
and
  \begin{equation}
  |c_1| + |c_2| < 2, \quad \omega_1 \neq \omega_2,
  \end{equation}
which is preferred by lasso, and the result is observed as two
signals at $\omega_1$ and $\omega_2$ around the expected signal
$\omega$.  For a numerical example, a lasso application to
the generated data for $t = 1, \cdots, 100$ and $\omega = 1/10$
in equation (\ref{equ:sinandcos}) yields $\omega_1 = 1/9.818$ and
$\omega_2 = 1/10.193$ for any $\lambda$ in 
$-2.58 < \log \lambda < -0.21$.\footnote{
   This split occurs regardless of the number of the data and
   length in time.  Although the split is smaller with longer length
   in time, the deviation from the expected signal is larger than
   the accuracy of period determination with the PDM.
}
  The coefficients $c_1 = 0.7632$ and
$c_2 = 0.7545$ yield $\frac{1}{2N} ||\bm{y} - A\bm{x}||_2^2 = 0.0039$
and $||\bm{x}||_1 = 1.5177$ in equation (\ref{equ:lassodef}).\footnote{
   These $c_1$ and $c_2$ were chosen to minimize
   $\frac{1}{2N} ||\bm{y} - A\bm{x}||_2^2$ in order to best
   illustrate the problem.  The actual values of $c_1$ and $c_2$
   selected by lasso are dependent on $\lambda$.  For example,
   $\lambda = 0.095$ yields $c_1 = 0.5488$ and $c_2 = 0.5599$,
   giving $\frac{1}{2N} ||\bm{y} - A\bm{x}||_2^2 = 0.0399$ and
   $||\bm{x}||_1 = 1.1087$.  In this case, $F(\bm{x}) = 0.145$
   and is indeed smaller than $F(\bm{x}) = 0.190$ for
   $\omega_1 = \omega_2 = 1/10$ and $c_1 = c_2 = 1$.
}
For $\omega_1 = \omega_2 = 1/10$ and $c_1 = c_2 = 1$,
the corresponding values are 0 and 2, respectively.
For $\lambda$ larger than 0.008, the former set of parameters
(with split signals) gives a smaller
$\frac{1}{2N} ||\bm{y} - A\bm{x}||_2^2 + ||\bm{x}||_1$.
Although a very small $\lambda$ might improve the situation,
such a small $\lambda$ will spoil the advantage of 
$\ell_1$ regularization.

   Although this difficulty can be avoided by introducing
complex coefficients and $\exp (-i\omega t)$-type basis
(\cite{can06exactsignal}; \cite{rau07harmonic}), this type
of $\ell_1$ regularization cannot be achieved by
{\bf lars} package.  As an alternative, practical approach to reduce
this false splitting of signals, we have
introduced an $N_k M \times N$ matrix in the following form:
  \begin{equation}\label{equ:fourierandphase}
  A_{k M+i,j} = \cos (\omega_i t_j + \frac{k}{N_k}\pi),
  \end{equation}
where $N_k$ is the number of phase bins and $k = 0, \cdots, N_k-1$.
Although the equation (\ref{equ:fourierandphase})-type basis is not
mathematically independent (it is an ``over-complete'' basis or
a ``frame''), the non-linear response of lasso,
caused by the characteristics of $\ell_1$-norm, enables us
to detect an adequate frequency and phase at the same time.
The number $N_k$ can be determined experimentally, and the
formulation becomes equivalent to equation (\ref{equ:fouriercomp})
in the case of $N_k = 2$.  In many of randomly sampled and randomly
phased actual data, there is no special need to use $N_k > 2$.
Even in most extreme cases, we have found that $N_k = 40$
successfully reproduced the original spectrum.

\begin{figure}
  \begin{center}
    \FigureFile(60mm,120mm){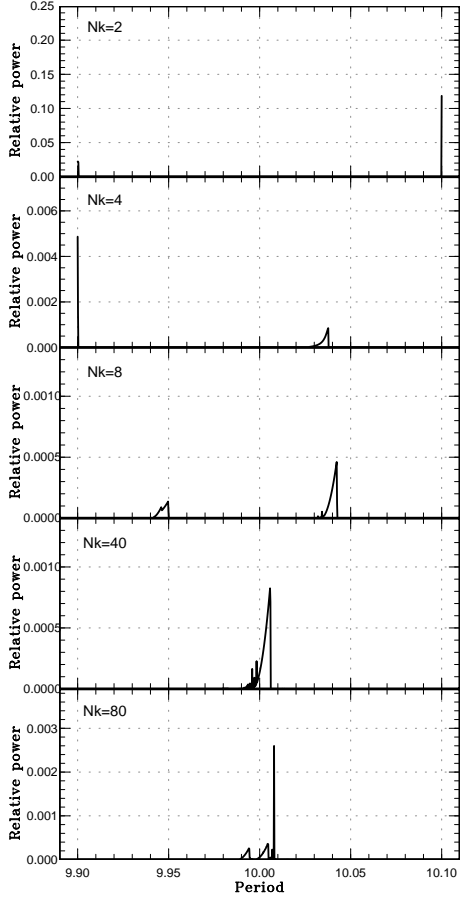}
  \end{center}
  \caption{Effect of $N_k$ on badly phased data.}
  \label{fig:nk}
\end{figure}

  We used an artificial data $\bm{x} = (1, \cdots, 100)^{\mathrm{T}}$
and used a function of $y(x) = \sin(2\pi x/10 + 0.63)$ to
produce $\bm{y}$.  The phase offset of 0.63 was chosen to
produce the most unfavorable condition.  The frequency window
corresponds to periods of 9.9--10.1.
The effect of different $N_k$ on this artificial data is shown in
figure \ref{fig:nk}.  Since our present aim is to determine
the single period precisely, we did not employ cross-validation,
which increases the number of detected signals, but used
a fixed $\log \lambda = -2.6$.
In the case of $N_k = 2$, the reproduced signals are located
on the upper and lower limits of the tested range.
In the case of $N_k = 4$, the separation between split peaks became
smaller, and we could obtain a merged signal with $N_k = 40$.
With an even larger $N_k$, false peaks were more prone to appear.

   Using $y(x) = \sin(2\pi x/10 + \pi/4)$, which corresponds
to the example in equation \ref{equ:sinandcos}, we could
obtain a reasonable result (single signal) with $N_k = 4$.
The experiment suggests that false splitting of signals is
expected to be avoided with $N_k$ up to 40, even in the worst cases.

\section{Conclusion}

   We introduced lasso in obtaining periodic signals in unevenly
spaced time-series data.  A very simple formulation with a combination
of a large set of sine and cosine functions has been shown to work
very well, and the peaks in the resultant power spectra were very sharp.
We studied the response of lasso to low signal-to-noise data,
asymmetric signals and very closely separated multiple signals.
When the length of the observation is sufficiently long, all of them
were not serious obstacles to lasso.  We analyzed the 100-year visual
observations of $\delta$ Cep, and obtained a very accurate period
of 5.366326(16) d.
The error in period estimation was several times smaller than in PDM.
We also modeled the historical data of R Sct, and obtained a reasonable
fit to the data.  The main two periods (143.8 d and 70.9 d) were not
exactly in a 2:1 ratio.  The model, however, lost its predictive ability
soon after the end of the data used for modeling, which is probably
a result of chaotic nature of the pulsations of this star.
We also formulated a scheme by using different set of phases in
the basis, and confirmed that this scheme is useful when
the phases of observed data are most unfavorable for lasso.  
\medskip

This work was supported by the Grant-in-Aid for the Global COE Program
``The Next Generation of Physics, Spun from Universality and Emergence"
from the Ministry of Education, Culture, Sports, Science and Technology
(MEXT) of Japan.
We are grateful to Prof. Masato Okada for introducing us to
the field of Compressed Sensing and to anonymous referees for
providing valuable suggestions and the literature.
We are grateful to VSOLJ and VSNET observers who contributed
observations we used.

\appendix

\section*{Sample R code}

   We provide a sample R code to make lasso analysis of time-series
data.  We assume that the data $\bm{t} = (t_1, \cdots, t_N)^{\mathrm{T}}$,
$\bm{y}$ are stored as two elements \verb|V1| and \verb|V2| in
the data frame \verb|d|.
The variable \verb|p| is a vector of assumed periods (for convenience,
we provide a function \verb|seqfreq| to make a series of periods evenly
spaced in the frequency space).  The lasso estimate can be obtained
by \verb|pow <- perlasso(d,p)|.  The parameter \verb|ndiv| corresponds
to $N_k$ in the text, and the parameter \verb|cv| is a flag whether
to compute cross-validations.  The results can be plotted by
\verb|plot(pow,n)|, where \verb|n| represents the $\lambda$ bin
(the values of $\lambda$ are stored in \verb|pow$lambda|).
The result of cross-validation is
stored in the list element of \verb|gcv| and the element \verb|nmin| is
the $\lambda$ bin giving the smallest MSE.
The function \verb|minper| returns the strongest signals and the function
\verb|lassofit| presents a fit to the data.

{\small
\begin{verbatim}
library(lars)
library(glmnet)

seqfreq <- function(a,b,...) {
    return(1/seq(1/b,1/a,...))
}

makematlasso <- function(d,p,ndiv) {
    nd <- length(p)
    m <- matrix(0,nrow(d),nd*ndiv)
    for (i in 1:nd) {
        ph <- ((d$V1/p[i]) %% 1)*pi*2
        for (j in 0:(ndiv-1)) {
            m[,i+nd*j] <- sin(ph+pi*j/ndiv)
        }
    }
    return(m)
}

perlasso <- function(d,p,ndiv=2,alpha=1,
            cv=FALSE) {
    nd <- length(p)
    mat <- makematlasso(d,p,ndiv)
    y <- d$V2 - mean(d$V2)
    m <- glmnet(mat,y,alpha=alpha)
    ndim <- m$dim[2]
    pow <- matrix(0,nd,ndim)
    for (i in 1:ndim) {
        v <- m$beta[,i]
        for (j in 0:(ndiv-1)) {
            pow[,i] <- pow[,i] +
               v[(nd*j+1):(nd*(j+1))]^2
        }
    }
    nmin <- NULL
    gcv <- NULL
    if (cv) {
        gcv <- cv.glmnet(mat,d$V2,alpha=alpha)
        minl <- gcv$lambda.min
        nmin <- which.min(abs(
                m$lambda-gcv$lambda.min))
    }
    r <- list(pow=pow,p=p,lambda=m$lambda,
         nmin=nmin,m=m,gcv=gcv,mat=mat)
    class(r) <- c("lassopow",class(r))
    return(r)
}

plot.lassopow <- function(pow,n,...) {
    p <- pow$p
    pw <- pow$pow
    plot(p,pw[,n],typ="l",xlab="Period",
         ylab="Relative Power",...)
}

minper <- function(pow,n,num=1) {
    p <- pow$p
    pw <- pow$pow
    pp <- numeric()
    while (num > 0) {
        i <- which.max(pw[,n])
        pp <- c(pp,p[i])
        num <- num-1
        pw[i,n] <- 0
    }
    return(pp)
}

lassofit <- function(pow,x,n,ndiv=2) {
    d <- data.frame(V1=x,V2=numeric(length(x)))
    m <- pow$mat
    return(m %*% pow$m$beta[,n])
}

lassoexpect <- function(pow,x,n,ndiv=2) {
    d <- data.frame(V1=x,V2=numeric(length(x)))
    m <- makematlasso(d,pow$p,ndiv)
    return(m %*% pow$m$beta[,n])
}
\end{verbatim}
}

\end{document}